\documentclass[preprint,showpacs,preprintnumbers,amsmath,amssymb,
superscriptaddress]{revtex4}

\usepackage{graphicx}
\usepackage{dcolumn}
\usepackage{bm}

\def\Bino{\tilde{B}}
\def\stau{\tilde{\tau}}
\def\stop{\tilde{t}}
\def\sfermion{\tilde{f}}

\def\stop{\tilde{t}}

\begin{document}

\preprint{KEK-TH-1017}
\preprint{STUPP-05-181}
\preprint{TUM-HEP-587/05}

\title{
Enhancement of Line Gamma Ray Signature from Bino-like Dark Matter Annihilation 
due to CP Violation
}

\author{Shigeki Matsumoto}
\affiliation{Theory Group, KEK, Oho 1-1, Tsukuba, Ibaraki 305-0801, Japan}
\author{Joe Sato}
\affiliation{Department of Physics, Saitama University, Saitama 338-8570, Japan}
\affiliation{Physik-Department, Techniche Universit\"at M\"unchen, 
James-Franck-Strasse D-85748 Garching, Germany}
\author{Yoshio Sato$^{2,3}$}
\date{\today}

\begin{abstract}
We study the line gamma ray signature from the Galactic Center 
originating from Bino-like dark matter pair annihilation, when CP is not conserved. 
We consider the case that the Bino is nearly degenerate with a sfermion in mass.
Although an indirect detection of Bino-like dark matter is difficult in general,
the signal of Bino-like dark matter pair annihialtion in this case is enormously enhanced
due to the threshold singularity.
\end{abstract}


\maketitle


Recent observations of cosmological and astrophysical quantities
determine the mean density of matter and baryons in the universe
precisely, and the existence of non-baryonic dark matter has now
been established\cite{WMAP}. 
However the composition of the dark matter is still
an unresolved problem. Weakly interacting massive particles (WIMPs) are
considered to be a prominent candidate for dark matter\cite{reviews}. 

The minimal supersymmetric standard model (MSSM) provides the stable 
lightest supersymmetric particle (LSP), if R-parity is conserved.
The neutralino LSP can be the dark matter due to the nature of the WIMP\cite{origin}.
The neutralino is a linear combination of the supersymmetric partners of
the U(1) and SU(2) gauge bosons (Bino and Wino), and those of the Higgs bosons 
(Higgsinos).

In many breaking scenarios, the Bino-like neutralino ($\Bino$) is predicted as the LSP. 
The viable models with bino-like dark matter require the
presence of mechanisms which suppress the density. 
One of the mechanism is the coannihilation\cite{coannihilation}. 
It works when there are other particles whose masses are nearly 
degenerate with that of the dark matter.
In this case, the relic abundance of the dark matter is
determined not only by its own annihilation but also by that of the slightly
heavier particles, which will later decay into the dark matter. 
For the Bino-like dark matter, the candidate for the coannihilating particle is
the sfermion $\sfermion$, which is the supersymmetric partner of the
fermion $f$. One such particleis the stau $\stau$, and the
coannihilation between Bino and stau has frequently been discussed in the past works, 
especially in the minimal supergravity model (MSUGRA). 
Another example is the stop $\stop$. This possibility is considered in the
context of electroweak baryogenesis in MSSM. 
The existence of the Bino-like dark matter having ${\cal O}(100)$ GeV mass 
is predicted in both cases.

Futhermore, the effects of CP violation have been considered not only on the
flavor physics but also on the dark matter physics.
In dark matter physics, the sfermion sector is extensively studied\cite{CP-violation}. 

For investigating the nature of the dark matter, many ideas for its
detection are proposed and some of the experiments are now operating. 
Among those, the detections of exotic cosmic ray fluxes\cite{BUB}, 
such as gamma rays, positrons, anti-protons and neutrinos from dark matter
annihilation are feasible techniques. In particular, an excess of
monochromatic (line) gamma rays originating from the dark matter pair annihilation
to two gammas would be a robust evidence if observed, because diffused
gamma ray background must have a continuous energy spectrum.
The satellite detectors and 
the large Atmospheric Cerenkov Telescope (ACT) arrays
may search for the exotic gamma rays from the Galactic Center\cite{telescope}.

In this letter, we study the line gamma ray flux from the annihilation
of the Bino-like dark matter which is nearly degenerate with stau or stop in mass
and has CP violating interaction with those particles. 
Interaction of the Bino-like neutralino is so weak that the line gamma ray flux 
from the annihilation has been considered to be very small without CP violation. 
We find, however, that the annihilation cross section is
enormously enhanced due to the threshold singularity when CP is
violated compared to that of the CP conserved case.
Since the dark matter is non-relativistic (the typical velocity is
${\cal O}(10^{-3})$), the relevant state for the incident Bino pair is
$^1S_0$, which is CP odd due to the Majorana nature, while the
sfermion pair in the $^1S_0$ state is CP even due to the bosonic nature. 
When CP is not conserved, the transition from the incident Bino pair to
a sfermion pair in the $^1S_0$ state takes place without the velocity suppression $v$
in the non-relativistic limit, and the sfermion pair can annihilate into two gammas. 


The source of the CP violation is contained in the off-diagonal element of 
the mass matrix for the third generation sfermion $M^2_{\sfermion}$, 
\begin{widetext}
\begin{eqnarray}
 M^2_{\sfermion}
 &=&
 \begin{pmatrix}
  M_L^2 + m_f^2 + m_Z^2 \cos{2\beta}~(T_{3L} - Q \sin^2{\theta_W})~
  &
  m^2_{\sfermion LR}
  \\
  m^{2*}_{\tilde{f}LR}
  &
  M_R^2 + m_f^2 + m_Z^2 \cos{2\beta}~Q\sin^2{\theta_W}
 \end{pmatrix}~,
\end{eqnarray}
\end{widetext}
where
\begin{eqnarray}
 m^2_{\sfermion LR}
 &=&
 -m_\tau(A^*_\tau + \mu \tan\beta)~,
 \qquad
 ({\rm for~stau})~,
 \nonumber \\
 &=&
 -m_t(A^*_t + \mu \cot\beta)~,
 \qquad~
 ({\rm for~stop})~,
\end{eqnarray}
$M_L$ and $M_R$ are the left- and right-handed sfermion masses, 
and $\tan\beta$ is the ratio of the vacuum expectation values of two Higgs fields. 
The charge and isospin of the sfermion are denoted by $Q$ and $T_{3L}$, respectively, 
and $\theta_W$ is the Weinberg angle. 
The parameters $m_f$, $m_Z$ and $\mu$ are the fermion, 
Z boson and supersymmetric Higgsino mass, respectively. 
The mass matrix of stau or stop is least studied experimentally, 
in particular the trilinear $A_\tau$ or $A_t$ parameter is not constrained from
stringent EDM (Electric Dipole Moment) experiments. 
Thus, there can be a wide space in $A_\tau$ or $A_t$ to accommodate large CP violation.

For following discussions, we define the sfermion mixing matrix for
diagonalizing the mass matrix, $U^\dagger M^2_{\sfermion} U = {\rm
diag.}(m^2_{{\sfermion}_1}, m^2_{{\sfermion}_2})$, as 
\begin{eqnarray}
 \begin{pmatrix}
  \sfermion_L
  \\
  \sfermion_R
 \end{pmatrix}
 =
 U
 \begin{pmatrix}
  \sfermion_1
  \\
  \sfermion_2
 \end{pmatrix}
 =
 \begin{pmatrix}
  \cos\theta_f & \sin\theta_f~e^{i\gamma_f}
  \\
  -\sin\theta_f~e^{-i\gamma_f} & \cos\theta_f
 \end{pmatrix}
 \begin{pmatrix}
  \sfermion_1
  \\
  \sfermion_2
 \end{pmatrix}~,
\end{eqnarray}
where we use the notation that $\sfermion_1$ has smaller mass than
$\sfermion_2$, $m_{\sfermion_1} \leq m_{\sfermion_2}$. Hereafter,
$\sfermion_1$ is denoted by $\sfermion$ and its mass  by $m_{\sfermion}$
for simplicity. We assume that the lightest sfermion $\sfermion$ is
almost degenerate with $\Bino$ in mass, which is needed for
the coannihilation process.
We also assume that other SUSY particles are so heavy, 
in general much heavier than $\sfermion$, that all the experimental
constraints are satisfied in the presence of the large $A_\tau$ or $A_t$.

The transition from the incident Bino pair to the sfermion pair is
caused by the Yukawa interaction $\Bino-\sfermion-f$.
Its strength is given by the gauge coupling due to supersymmetry,
\begin{eqnarray}
 {\cal L}_{tr}
 &=&
 \sfermion^*\bar{\Bino}
 \left(
  a_L P_L
  +
 a_R P_R
 \right)f
 +
 {\rm h.c.}~,
 \label{CP non-conserving}
\end{eqnarray}
where,
\begin{eqnarray}
 a_L
 &=&
 -\sqrt{2}g'(Q - T_{3L})\cos\theta_f
 ~, \\
 a_R
 &=&
 -\sqrt{2}g'Q\sin\theta_f~e^{i\gamma_f}
 ~,
\end{eqnarray}
$g'$ is the U(1)$_Y$ gauge coupling constant of the standard model.
The effect of CP violation is induced by 
the relative complex phase $e^{i\gamma_f}$.

\begin{figure}[h]
 \begin{center}
  \includegraphics[width=70mm,keepaspectratio,clip]{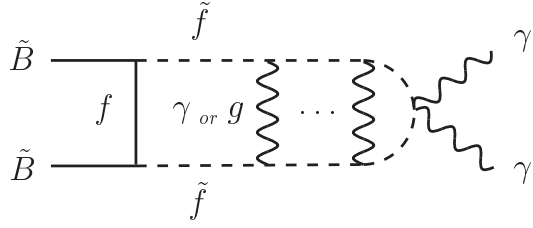}
  \caption{\small 
  Ladder diagram contributing to the threshold singularity in
  $\Bino\Bino\rightarrow\gamma\gamma$ process. The character $g$ means
  gluon, which is important when $f$ is top. }
 \label{diagram}
 \end{center}
\end{figure}

The threshold singularity appears in the calculation of the
annihilation cross section due to the degeneracy in mass between a Bino pair
and an intermediate sfermion pair and their non-relativistic motions.
The intermediate sfermion pair forms a quasi stable bound state
when the binding energy is approximately equal to the mass difference,
and hence the annihilation cross section is enhanced.
In a Feynman diagrammatic picture, higher order terms such as a
ladder diagram depicted in Fig.\ref{diagram}, significantly contribute to
the cross section, because an intermediate sfermion pair is almost on-shell. 
Thus, we have to resum the ladder diagram for a reliable calculation. 
The non-relativistic (NR) effective theory is useful to treat the singularity\cite{NR-L},
which is often used in calculations of the threshold production of
particles, the quarkonium mass spectrum and so on. 
The Schr\"odinger equation for the non-relativistic Bino pair annihilation
is derived from the NR effective Lagrangian as its equation of motion. 
Solving the Schr\"odinger equation 
is equivalent to the resummation of ladder diagrams in Fig.\ref{diagram}.

The derivation of the Lagrangian here is essentially the same as that 
in Ref.~\cite{non-perturbative}. Therefore we show only the result. 
The NR effective Lagrangian for two-body states,
$\phi_{\Bino}(\vec{r})~(\simeq \Bino\Bino/2)$ and 
$\phi_{\sfermion}(\vec{r})~(\simeq \sfermion^*\sfermion)$, is given as 
\begin{eqnarray}
 {\cal L}
 =
 {\bf \Phi}^\dagger(\vec{r})
 \left[
  \left(
   E + \frac{\nabla^2}{m}
  \right){\bf 1}
  -
  {\bf V}(\vec{r})
  +
  iC{\bf \Gamma}\delta^3(\vec{r})
 \right]
 {\bf \Phi}(\vec{r})~,
 \label{NRL}
\end{eqnarray}
where ${\bf \Phi}^\dagger(\vec{r}) = (\phi^*_{\sfermion}(\vec{r}),
\phi^*_{\Bino}(\vec{r}))$, $\vec{r}$ is the relative coordinate ($r =
|\vec{r}|$), $E$ is the internal energy of the two-body state$E=mv^2/4$, 
$m$ is the Bino mass, $v$ is the relative velocity of 
the incident dark matter pair ($v=|\vec{v}|$).
The parameter $C$ is the color factor, which takes a
value $C = 1$ for stau and $C = 3$ for stop. The potential ${\bf V}(r)$
has $2\times 2$ matrix form. The (1,1) component describes the force
acting between the sfermion pair and the dominant contribution is the
coulomb force for the stau case, while it comes from the QCD interaction
for the stop case,
\begin{eqnarray}
 {\bf V}_{11}(r)
 &=&
 2\delta m
 -
 \frac{d}{r}~,
 \label{potential-11}
\end{eqnarray}
where $\delta m$ is the mass difference between the sfermion and Bino,
$d=\alpha$ for the stau case, and $d=4\alpha_s/3$ for the stop case.
We set the fine structure constant $\alpha=1/128$, and 
the strong coupling $\alpha_s$ is determined by solving the one-loop renormalization group equation
at the energy $4m\alpha_s/3$;
\begin{eqnarray}
 \alpha_s(4mx/3) = x~\,
 \label{consistency}
\end{eqnarray}
where $\alpha_s(\mu)$ is the effective coupling at scale $\mu$. 
Eq.~\eqref{consistency} is derived from the consistency equation that
the typical momentum of the intermidiate gluon in the t-channel is about $md$,
when the sfermion pair forms a bound state by the attractive force of 
the second term of right hand side in ~\eqref{potential-11}.
The (1,2) (= (2,1)) component represents the
transition between the sfermion pair and the Bino pair, and is derived from
the interaction in eq.(\ref{CP non-conserving}),
\begin{eqnarray}
 {\bf V}_{12}(r)
 =
 \frac{f}{r}e^{-m_f r}~,~
 \label{off-diagonal}
\end{eqnarray}
where
\begin{eqnarray}
 f= \frac{\sqrt{C} m_f}{4\pi m} \Im (a_R a_L^*)~~\propto~\sin 2\theta_f \sin\gamma_f~.~
\end{eqnarray}
Since the transition takes place due to the CP-violating interaction, 
the component is proportional to $\sin\gamma_f$ as expected. 
The (2,2) component of the potential ${\bf V}$ is zero, 
because there are no forces acting between Bino pair in the non-relativistic limit. 
The imaginary (absorptive) part of the effective Lagrangian ${\bf \Gamma}$ is
\begin{eqnarray}
 {\bf \Gamma}_{ij}
 =
 \frac{2\pi Q^4\alpha^2}{m^2}\delta_{i1}\delta_{j1}~.
\end{eqnarray}
The sfermion pair annihilates into two gammas through this part.

\begin{figure}[h]
  \begin{center}
    \includegraphics[width=60mm]{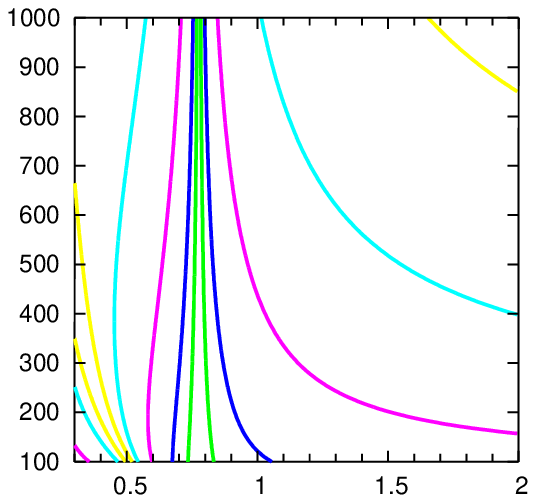}
    \put(-190,175){(a)}  
    \put(-125,165){$\sigma v$ (10$^{-29}$ cm$^3$ s$^{-1}$)}
    \put(-185,65){\rotatebox{90}{$m$ (GeV)}}
    \put(-105,-12){$\delta m/m$ (10$^{-5}$)}
    \put(-25,125){\scriptsize $0.01$}
    \put(-40,80){\scriptsize $0.1$}
    \put(-70,45){\scriptsize $1$}
    \put(-92,25){\scriptsize $10$}
    \put(-110,17){\scriptsize $100$}
    \put(-123,30){\tiny $10$}
    \put(-125,75){\scriptsize $1$}
    \put(-140,130){\scriptsize $0.1$}
    \put(-152,50){\scriptsize $0.01$}
  \hspace{20mm}
    \includegraphics[width=60mm]{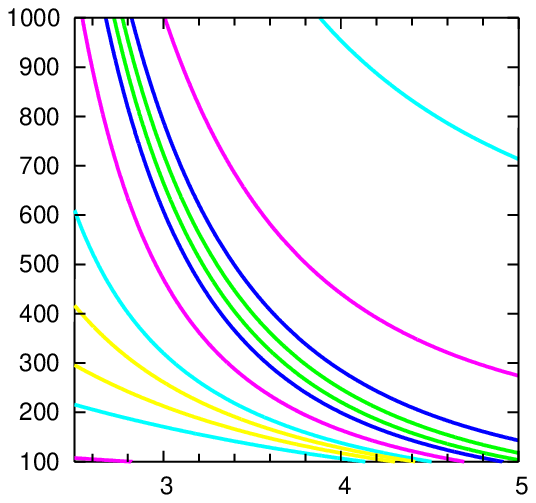}
    \put(-190,175){(b)}
    \put(-125,165){$\sigma v$ (10$^{-29}$ cm$^3$ s$^{-1}$)}
    \put(-185,65){\rotatebox{90}{$m$ (GeV)}}
    \put(-105,-12){$\delta m/m$ (10$^{-3}$)}
    \put(-30,125){\scriptsize $0.01$}
    \put(-45,59){\scriptsize $0.1$}
    \put(-51,38){\scriptsize $1$}  
    \put(-90,50){\scriptsize $10$}
    \put(-133,113){\scriptsize $1$}
    \put(-141,90){\scriptsize $0.1$}
    \put(-139,60){\scriptsize $0.01$}
    \put(-145,45){\scriptsize $0.001$}
    \put(-140,19){\scriptsize $0.01$}
    \vspace{0mm}
    \caption{{\footnotesize Contour plot of the annihilation cross section of Bino-like dark matter to two gammas.
             (a) is stau case, and (b) is stop case.
             We take the sfermion mixing as $\theta_f=\pi/4$, and CP violating phase as $\gamma_f=\pi/2$.}}
  \label{cross-section}
  \end{center}
\end{figure}

The amplitude for the Bino dark matter pair annihilation is obtained by solving the
Schr\"odinger equation 
\begin{eqnarray}
 \left[
  -\frac{\nabla^2}{m}
  +
  {\bf V}(\vec{r})
 \right]{\bf \Phi}(\vec{r})
 =
 E{\bf \Phi}(\vec{r})~,
 \label{Schroedinger eq}
\end{eqnarray}
where the boundary condition is given by
\begin{eqnarray}
 {\bf \Phi}(\vec{r} \to \infty)
 =
 \left(
  \begin{array}{@{\,}c@{\,}}
  0 \\ e^{i\vec{k}\cdot\vec{r}}
  \end{array}
 \right),~~
 \vec{k}=\frac{m}{2}\vec{v}~.
 \label{bc}
\end{eqnarray}
The annihilation cross section of Bino-like dark matter to two gammas is calculated as
\begin{eqnarray}
 \sigma v
 = \left< {\bf \Phi}| C {\bf \Gamma}| {\bf \Phi} \right>
 = \frac{2\pi C Q^4 \alpha^2 }{m^2} |\phi_{\tilde{f}}(0)|^2~. \label{crossec}
\end{eqnarray}
We perturbatively expand the solution of eq.~\eqref{Schroedinger eq} 
by the off-diagonal element of the potential~\eqref{off-diagonal}.
Under the boundary condition ~\eqref{bc}, 
the solution to 1st order in the perturbation is given by
\begin{eqnarray}
 \phi_{\tilde{f}}(0)
&=&
 -\frac{2f}{v}\int_0^\infty dy~
  \frac{e^{-m_f y/dm}}{y} t(y) \sin{\left(\frac{vy}{2d}\right)}~.\label{1st-solution}
\end{eqnarray}
where,
\begin{eqnarray}
 t(x) =
 -\Gamma\left(-\frac{1}{2\sqrt{a}}\right)e^{-\sqrt{a}x}~x~U(1-\frac{1}{2\sqrt{a}},2,2\sqrt{2}x)~,
\end{eqnarray}
where $a=\frac{1}{d^2m}(2\delta m - \frac{m v^2}{4}) $, 
$\Gamma(x)$ is Euler's gamma function,
and the function $U(a,b,c)$ is the confluent hyper geometric function defined by
\begin{eqnarray}
 U(a,b,c) \equiv \frac{1}{\Gamma(a)}\int_0^\infty dt~ e^{-ct}~t^{a-1}~(1+t)^{b-a-1}~.
\end{eqnarray}
Then the annihilation cross section of Bino-like dark matter is given by
substituting~\eqref{1st-solution} for~\eqref{crossec}.
In Fig.~\ref{cross-section}, the numerical results for the annihilation cross
section to two gammas are shown.
We assume that the sfermion mixing is maximal and CP violating phase is also maximal.
Our result is sensitive to the sfermion mixing and CP violating phase.
However this result is maintained as long as the sfermion mixing is not so small and the CP violating phase is ${\cal O}(1)$.

The annihilation cross section is enormously enhanced, 
if $\delta m$ is approximately equal to the binding energy of the sfermion pair. 
The binding energy is determined by 
\begin{eqnarray}
 \frac{m d^2}{4n^2} = 2\delta m - \frac{mv^2}{4}~,
\end{eqnarray}
where $n$ is a principal quantum number.
When $n=1$, $\delta m/m \simeq d^2/8$.
Indeed the enhancement is realized there as shown in Fig.~\ref{cross-section}.

We note that the neutralino annihilation cross section to two gammas
has been calculated in the CP conserving case\cite{loop},
and it rarely exceeds a few times $10^{-30}$ cm$^3$s$^{-1}$ for Bino-like neutralinos \cite{BUB}.
On the contrary, in our case, the annihilation cross section
can be enhanced by several orders of magnitude compared to the CP conserving case.

The enhancement of the annihilation cross section gives significant
impact on an indirect detection for the dark matter search, because
the flux of line gamma rays is proportional to the cross section. 
We discuss the line gamma ray flux from the dark matter
annihilation in the Galactic Center, which is given by
\begin{eqnarray}
 F_{\rm line}
 &=&
 1.9\times 10^{-11}
 \left(\frac{100~{\rm GeV}}{m}\right)^2
 \left(\frac{\langle \sigma v\rangle}{\rm 10^{-27}cm^3s^{-1}}\right)
 \bar{J}(\Delta\Omega)~\Delta\Omega~{\rm cm^{-2}s^{-1}} ~,
 \label{Jbar}
\end{eqnarray}
where $\left< \sigma v \right>$ is the thermally averaged annihilation cross section,
and $\Delta\Omega$ is the angular acceptance of the detector, 
taken to be $10^{-3}$ for our calculation, which is a typical value for ACT detectors. 
The flux depends on the dark matter profile through 
the parameter $\bar{J}(\Delta\Omega)$ in eq.(\ref{Jbar}),
\begin{eqnarray}
 \bar{J}
 \equiv
 \int_{\Delta\Omega}\frac{d\Omega}{\Delta\Omega}
 \int_{\rm line~of~site}\frac{dl}{8.5~{\rm kpc}}
 \left(\frac{\rho}{0.3~{\rm GeVcm^{-3}}}\right)^2,
\end{eqnarray}
where the integral over $l$ is performed along the line of sight. 
The parameter $\bar{J}(\Delta\Omega)$ is estimated in various halo model, and takes values
$10 < \bar{J}(10^{-3}) < 10^6$. 
We take $\bar{J}(10^{-3}) = 1352$ for the NFW profile\cite{NFW}, which is a moderate value\cite{reviews}.

\begin{figure}[h]
  \begin{center}
    \includegraphics[width=60mm,keepaspectratio,clip]{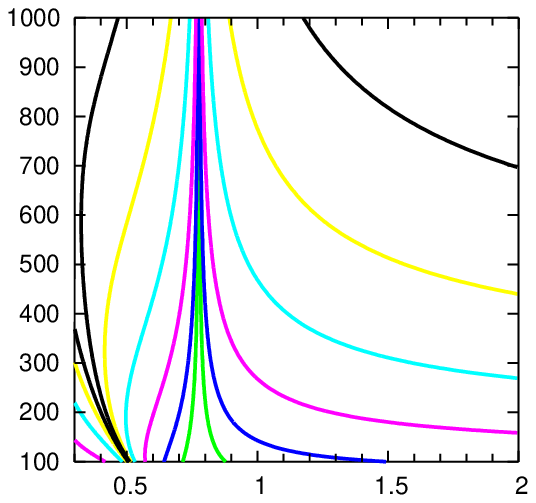}
    \put(-190,175){(a)}  
    \put(-125,165){$F_{\rm line}$ (10$^{-14}$ cm$^2$ s$^{-1}$)}
    \put(-185,65){\rotatebox{90}{$m$ (GeV)}}
    \put(-105,-12){$\delta m/m$ (10$^{-5}$)}
    \put(-25,117){\scriptsize $0.01$}
    \put(-45,80){\scriptsize $0.1$}
    \put(-60,52){\scriptsize $1$}
    \put(-75,33){\scriptsize $10$}
    \put(-87,20){\scriptsize $100$}
    \put(-110,17){\tiny $1000$}
    \put(-125,48){\tiny $10$}
    \put(-126,75){\scriptsize $1$}
    \put(-136,115){\scriptsize $0.1$}
    \hspace{20mm}
    \includegraphics[width=60mm,keepaspectratio,clip]{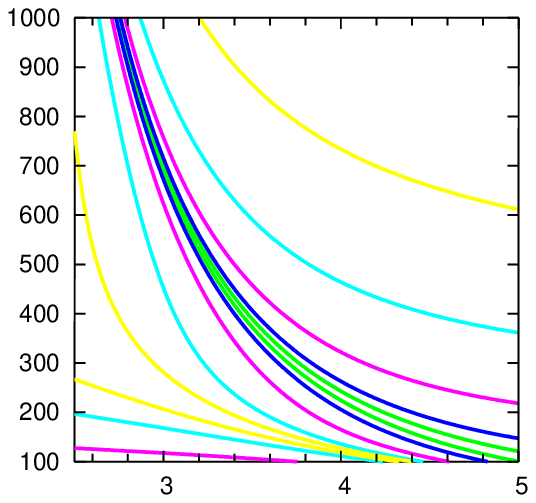}
    \put(-190,175){(b)}
    \put(-125,165){$F_{\rm line}$ (10$^{-14}$ cm$^2$ s$^{-1}$)}
    \put(-185,65){\rotatebox{90}{$m$ (GeV)}}
    \put(-105,-11){$\delta m/m$ (10$^{-3}$)}
    \put(-80,127){\scriptsize $0.01$}
    \put(-70,75){\scriptsize $0.1$}
    \put(-53,45){\scriptsize $1$}
    \put(-40,29){\scriptsize $10$}
    \put(-30,17){\tiny $100$}
    \put(-97,40){\scriptsize $1$}
    \put(-133,70){\scriptsize $0.1$}
    \put(-145,45){\scriptsize $0.01$}
    \vspace{0mm}
    \caption{{\footnotesize Contour plot of the line gamma ray flux from the Galactic Center 
             originating from Bino-like dark matter pair annihialtion.
             (a) is stau case, and (b) is stop case.
             We take $\Delta\Omega=10^{-3}$, $\bar{J}(10^{-3})=1352$, $\theta_f=\pi/4$,
             and $\gamma_f=\pi/2$.}}
    \label{flux}
  \end{center}
\end{figure}

Our prediction for the line gamma ray flux from the Galactic Center is shown in Fig.~\ref{flux}.
The line gamma ray flux is enourmously enhanced, and will be observed by the ACT detectors. 

To conclude,
we have calculated the line gamma ray flux from the Galactic Center originating from
the Bino-like dark matter annihilation in the CP violating case.
We have showed that the line gamma ray flux is enourmously enhanced due to the threshold singularity
when the degeneracy between Bino and sfermion is sufficiently small.

HESS and CANGAROO-II have already taken data from the Galactic Center.
Some parameter region may have been excluded by these ACT detectors.
However, the results of HESS and CANGAROO-II show the discrepancy\cite{discrepancy},
and hence we do not compare our prediction with these results in detail.
Our predicted flux is large and hence the line gamma ray will be observed by the ACT detectors in the near future,
even if the neutralino is Bino-like.

\section*{Acknowledgments}
We thank M.M.Nojiri for useful discussion.
The work of J.S is supported by the Grant-in-Aid for
Scientific Research on Priority Area No.16028202 and 17740131.

\end{document}